# Evaluation of Analytical Models in Scattering Scanning Near-field Optical Microscopy for High Spatial Resolution Spectroscopy

Soheil Khajavi [1] soheil.khajavi@ut.ac.ir, Ali Eghrari [3] ali.eghrari@kcl.ac.uk, Zahra Shaterzadeh-Yazdi [2] zahra.shaterzadeh@ut.ac.ir, Mohammad Neshat [1] mneshat@ut.ac.ir

1. School of Electrical and Computer Engineering, College of Engineering, University of Tehran, Tehran, Iran
2. School of Engineering Science, College of Engineering, University of Tehran, Tehran, Iran
3. Faculty of Natural, Mathematical & Engineering Sciences, King's College London, London, United Kingdom

**Abstract-** Scattering scanning near-field optical microscopy (s-SNOM) is a technique to enhance the spatial resolution, and when combined by Fourier transform spectroscopy it can provide spectroscopic information with high spatial resolution. This paper studies two analytical models for the s-SNOM probe using atomic force microscopy (AFM) tip and its interaction with a dielectric material. We evaluate the validity of these models by retrieving the permittivity spectrum of a sample material through an inverse method.

## 1. Introduction

Various techniques have been introduced to overcome the diffraction limit in spatial resolution [1,4]. These techniques mostly use an aperture [5] or apertureless [6] probes, placed close to a sample to convert the evanescent waves to propagating waves. Scattering scanning near-field optical microscopy (s-SNOM) appeared to be a viable solution due to providing decoupling between spatial resolution and wavelength. In s-SNOM an atomic force microscopy (AFM) tip is normally used as a probe in which the resolution is determined by the size of the AFM tip regardless of the incident field's wavelength. As shown in Fig.1, the scattered field from the AFM tip that is in proximity of a sample under test, is coupled to a Michelson interferometer to extract the spectral properties of the scattered field.

In this paper, we review two types of dipole approximation to model AFM tip, and use the image theory to calculated the scattered field from an AFM tip located close to a dielectric sample. The accuracy of the models is evaluated by retrieving the permittivity of a sample material from the scattered fields.

In the following two theoretical models for AFM tip are analysed. Then after introducing an inverse method, they are evaluated by the resulting figures.

## 2. Analytical Models of Field Scattering from AFM-tip

In this section, we review two types of dipole approximation to model AFM tip, and use the image theory to calculated the scattered field from an AFM tip located close to a dielectric sample. The accuracy of the models are evaluated by retrieving the permittivity of samples from the scattered fields.

### 2.1. Point Dipole Approximation

The size of a tip is normally much smaller than the illuminating wavelength. In point dipole approximation, the AFM tip over a sample is modelled by a dielectric sphere with radius $a$ ($a << \lambda$) over a half-space dielectric as illustrated in Fig. 2. The distance from the centre of sphere to the surface of the half-space dielectric sample is $r$. The dielectric permittivity of the sphere and the sample are $\epsilon_1$ and $\epsilon$, respectively.

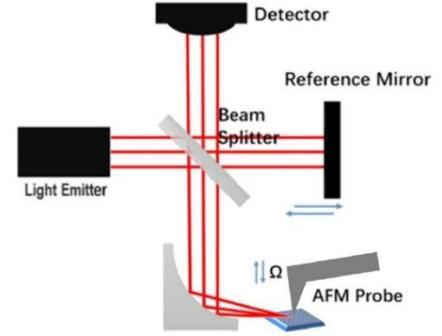

Fig. 1: A simplified s-SNOM with AFM probe combined with Michelson interferometer as a Fourier transform spectrometer with high spatial resolution.

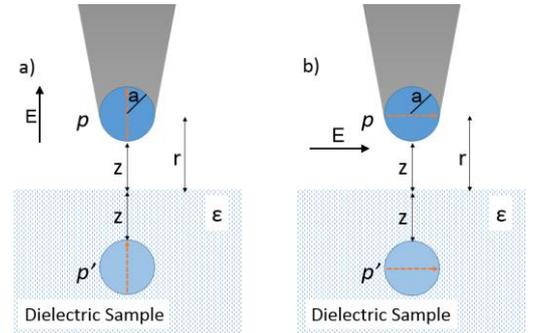

Fig. 2: Simplified geometrical model of AFM tip over a sample. The dielectric sphere is polarized by a) vertical, b) horizontal electric field.

Since the radius of the sphere in Fig. 2 is much smaller than the wavelength, one can estimate the polarizability, $\alpha$, and the induced dipole moment, $\vec{p}$, of the sphere through quasi-static approximation as [7]

$$\vec{p} = \alpha \vec{E}, \; \alpha = \left(\frac{\epsilon_1 - 1}{\epsilon_1 + 2}\right) \epsilon_0 a^3 \qquad (1)$$

Where $\vec{E}$ is the electric field at the sphere. The electric field, resulting from a dipole moment $p$ along its axis is obtained as

$$E_{\text{dipole}}(R) = \frac{p}{2\pi R^3} \qquad (2)$$

Where $R$ is the distance from the observation point to the point dipole. The field at the sphere is influenced by the sample underneath. Such influence can be accounted for by a dipole moment image $\vec{p'}$ located inside the dielectric half-space area, and at the same distance from the sample surface as that of the initial dipole (see Fig. 1). The image dipole moment can be obtained as

$$\vec{p'} = \beta\vec{p}, \quad \beta = \left(\frac{\epsilon-1}{\epsilon+1}\right) \tag{3}$$

To consider the effect of the sample on the induced dipole moment of the tip, the electric field at the sphere is assumed as the summation of the incident field and the generated field from the image dipole, resulting in a modified dipole moment for the tip as

$$p = \alpha\left(E + \frac{p'}{16\pi r^3}\right) = \alpha\left(E + \frac{\beta p}{16\pi r^3}\right) \tag{4}$$

Solving 4 for the value of $p$, i.e. the modified electric dipole moment of the tip, one obtains

$$p = \frac{\alpha}{1 - \frac{\alpha\beta}{16\pi r^3}} E \tag{5}$$

Therefore, the effective polarizability of the modified electric dipole moment for the vertically polarized tip is

$$\alpha_\perp^{\text{eff}} = \frac{\alpha}{1 - \frac{\alpha\beta}{16\pi r^3}} \tag{6}$$

The same procedure can be followed for horizontally polarized tip (Fig. 2b). The electric field, resulting from a dipole moment $p$ along a perpendicular direction to its axis is obtained as

$$E_{\text{dipole}}(R) = -\frac{p}{4\pi R^3} \tag{7}$$

The effective polarizability for the horizontal case is then obtained by

$$\alpha_\parallel^{\text{eff}} = \frac{\alpha(1-\beta)}{1 - \frac{\alpha\beta}{32\pi r^3}} \tag{8}$$

Equations (6) and (8) show that the effective polarizability contains the information about the permittivity of both sample and the tip. Moreover, for samples with large permittivity, the effective polarizability due to vertical fields dominates.

### 2.2. Finite Dipole Approximation

In finite dipole approximation, the AFM tip is modelled by a dielectric ellipsoid as shown in Fig. 3. It is polarized by the incident field $E_{in}$. The induced charges in the ellipsoid due to the incident field is modeled by point charges $\pm Q_0$ at $W_0 \approx 1.31a$ from its apices, where $a$ is the radius of the fitted sphere to the tip [8]. The resultant induced dipole moment $p_0$ is approximated by

$$p_0 \approx 2LQ_0 \tag{9}$$

Where $2L$ is the ellipsoid length. The $+Q_0$ monopole due to its closeness to the sample can interact with it, resulting in an image with the charge $-\beta Q_0$, where $\beta$ is given in (3). This image charge induces another monopole inside the tip as $Q_1$ at $W_1 \approx a/2$ from the apex as shown in Fig. 3.

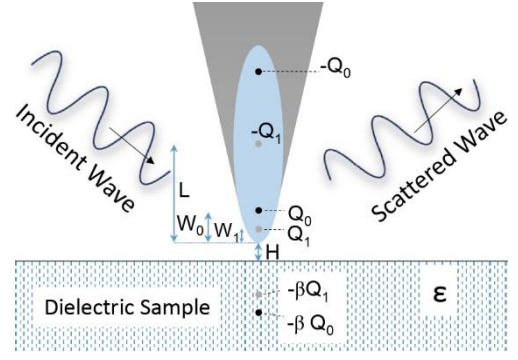

Fig. 3: Modeling of the AFM tip with a dielectric ellipsoid. The equivalent polarized charges due to the incident wave and the effect of sample is represented by point charges $\pm Q_{0,1}$.

To calculate the induced charge $Q_1$, a self-consistent treatment is used to map the effect of images of $Q_0$ and $Q_1$ on themselves [9]. So that $Q_1$ is related to $Q_0$ as

$$Q_1 = \frac{\beta f_0}{1 - \beta f_1} Q_0 \tag{10}$$

Where $f_0$ and $f_1$ are functions containing geometrical parameters illustrated in Fig. 3 and are given by [8]

$$f_{0,1} = \left(g - \frac{a + 2H + W_{0,1}}{2L}\right) \frac{\ln\frac{4L}{a + 4H + 2W_{0,1}}}{\ln\frac{4L}{a}} \tag{11}$$

Where $g$ is an empirical geometry factor that for typical AFM tips, $|g| = 0.7 \pm 0.1$ [10].

The charge $-Q_1$ is placed at distance $L$ from its positive counterpart in the ellipsoid to neutralize the whole charges, resulting in the dipole moment $p_1 \approx Q_1 L$. The scattered light from the tip is proportional to the effective polarizability of the tip.

$$\alpha^{\text{eff}} \propto 1 + \frac{p_1}{p_0} = 1 + \frac{1}{2}\frac{\beta f_0}{1 - \beta f_1} \tag{12}$$

In section IV, we use these two approximation models to retrieve the permittivity of samples through an inverse problem.

### 2.3. Oscillating AFM probe

In s-SNOM setups, as shown in Fig. 1, most of the received power by the detector is due to the scattered light from the background including the sample and the cantilever body. However, the useful information lies in the scattered light due to the near-field interaction of the tip with sample. Therefore, to separate the desired signal from the background, the tip mechanically oscillates by a specific frequency, $\Omega$. Therefore, the distance of the tip from the sample changes as

$$H(t) = H_0 + A\cos(\Omega t) \tag{13}$$

Where $H_0, A$ and $\Omega$ are the average distance, amplitude and frequency of the mechanical oscillation, respectively. Due to high dependency of the scattered wave to the near-field interaction of the tip with sample, the oscillation of the tip modulates the scattered wave by the mechanical oscillation frequency. Therefore, the desired signal can be demodulated at the oscillation frequency and its harmonics, and separated from the undesired background [11].

To retrieve the spectrum of the effective polarizability, $\alpha^{\text{eff}}$, the scattered field is coupled to a Michelson interferometer as shown in Fig.1. Then, the spectral information can be extracted from a standard Fourier analysis as used in conventional Fourier transform spectrometers.

## 3. An Inverse Method to Retrieve Sample Permittivity

An inverse method is used to retrieve the sample permittivity. For the sake of demonstration, we use point dipole approximation in the following formulations, however, the finite dipole approximation follows the same algorithm. The ratio of the scattered to the incident field is proportional to the effective polarizability as

$$\sigma = \frac{E_{\text{sc}}}{E_{\text{in}}} = \alpha_\perp^{\text{eff}} \quad (14)$$

Where $\alpha_\perp^{\text{eff}}$ is given in (6) and can be rewritten as $\alpha_\perp^{\text{eff}} = \frac{\alpha}{1-f\beta}$ with $f = \alpha/(16\pi r^3)$. For practical cases, the condition $|f\beta| < 1$ is satisfied. As a result, $\alpha_\perp^{\text{eff}}$ can be expanded into Taylor series with respect to $f\beta$ as

$$\alpha_\perp^{\text{eff}} = \sum_{j=0}^{\infty} a_j (f\beta)^j \quad (15)$$

Where $a_j$ is the Taylor series coefficient. It should be noted that parameter $f$ in (15) is a periodic function in time, because it is a function of the oscillating distance between the tip and the sample. Therefore, parameter $\sigma$ can be expanded into a Fourier series with harmonic coefficients as

$$\sigma_n = \sum_{j=1}^{\infty} \hat{F}_n\{a_j f^j\} \beta^j \quad (16)$$

Where $\sigma_n$ is the $n^{th}$ harmonic Fourier coefficient, $\hat{F}_n\{.\}$ is an operator extracting $n^{th}$ harmonic coefficient. In experiments, to calibrate out the frequency response of the tip, detector and all other optical devices, the measured signal harmonic, $\sigma_n$, from the unknown sample should be normalized to that of a reference material (with known $\beta$) as

$$\eta_n = \frac{\sigma_n}{\sigma_{n,ref}} = \frac{1}{\sigma_{n,ref}} \sum_{j=1}^{J} \hat{F}_n\{a_j f^j\} \beta^j \quad (17)$$

It should be noted that $\eta_n$ is calculated from the ratio of $n^{th}$ harmonics of two measured signals recorded for unknown and reference materials. Equation (17) gives rise to an $J$-order polynomial equation. By solving each parameter, $\beta$ is obtained over the desired frequency range. Once the spectrum of $\beta$ is known, the permittivity of the sample can be obtained from (3).

## 4. Evaluation of the Analytical Models

In this section, we evaluate the performance of the analytical models discussed in previous sections. First, we apply the forward method to calculate the spectrum of the scattered field for a PMMA polymer sample. Its permittivity spectrum was obtained from [12], and plotted over the wavelength range of 5-7 $\mu$m in Figs. 4a and b. The effective polarizability of the tip over the PMMA sample is compared for point and finite dipole approximations in Figs. 4c and d. The tip is made of platinum with a distance of 3 nm from the sample surface. In point dipole model, the radius of the sphere is 30 nm.

In finite dipole model, the length of the ellipsoid is considered 1.2 $\mu$m, and the radius of the sphere is 30 nm.

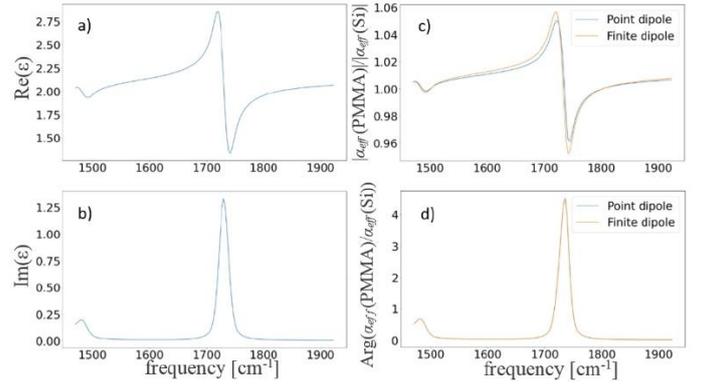

FIG. 4. (a) Real and (b) imaginary part of the permittivity of PMMA polymer for wavelength range of 5-7 $\mu$m[12]. Comparison of (c) amplitude and (d) phase of the effective polarizability of the tip over PMMA sample for point and finite dipole approximations. The phase plots are perfectly overlaid on one another in (d).

As it is observed in Fig. 4c and d, the two models provide quite similar results at distance 3 nm. As the distance between the tip and sample increases, differentiation in the results begins to appear as demonstrated in Fig. 5 for distances of 10 nm and 30 nm. As the distance increases the polarizability is more affected in point dipole model. The finite dipole method can also model the interaction with higher distance.

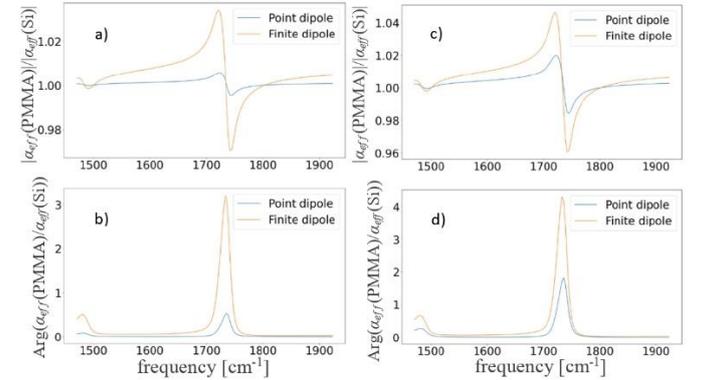

FIG. 5. Effective polarizability (a) amplitude and (b) phase for distance of 30 nm, (c) amplitude and (d) phase for distance of 10 nm.

In Fig. 6, the amplitude and phase of the normalized scattered field for the $3^{rd}$ harmonic, $\eta_3$, obtained from point and finite dipole models and the results from an experiment are compared. Tip oscillation with frequency of $\Omega = 350$kHz and amplitude of 30 nm is considered. The blue curve in Fig. 6 is the experimental result from [13], for PMMA which is normalized to calcium fluoride reference material. As it is observed in Fig. 6, the finite dipole model provides closer results to the experimental data.

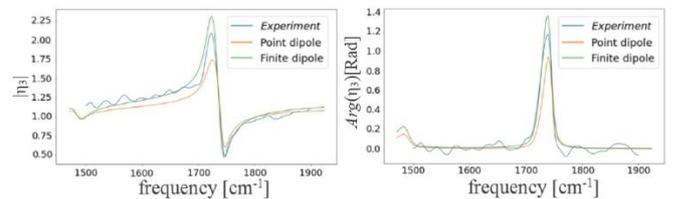

FIG. 6. Normalized amplitude and phase of the scattered field as a function of frequency for the $3^{rd}$ harmonic. The blue curves are experimental results from [13].

Figure 7 compares the retrieved permittivity spectrum of PMMA from the $3^{rd}$ harmonic through the inverse method by point dipole and finite dipole models with experimental ellipsometry results reported in [13]. We considered nine terms of Taylor series in the calculations. As it is observed in Fig 7, the finite dipole model provides much better prediction.

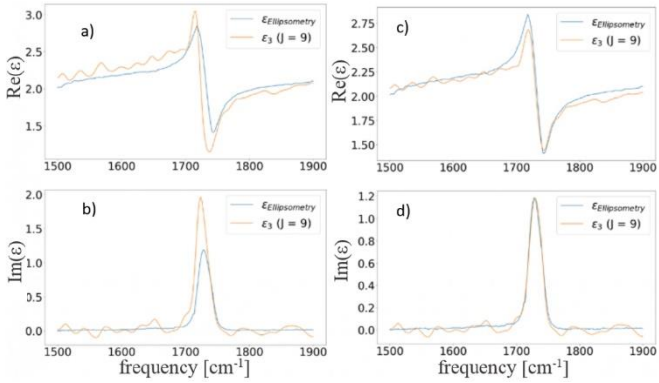

FIG. 7. Retrieved permittivity spectrum from the $3^{rd}$ harmonic through the inverse method by point dipole, a) real part, b) imaginary part, and finite dipole c) real, d) imaginary part. Graphs in blue are from the experimental ellipsometry measurement reported in [13].

## 5. Conclusion

To conclude, we evaluated, validated and compared two analytical approximations for the AFM probe for enhancing the spatial resolution of the spectroscopy setups. The results show that the finite dipole approximation seems to be more accurate and provides closer results to the experimental data, which can be used to determine the permittivity of more materials in the future research.